\documentclass[epj]{webofc}
\usepackage[utf8]{inputenc}
\usepackage[varg]{txfonts}   
\usepackage{booktabs}
\usepackage{xcolor}
\definecolor{darkred}{rgb}{0.4,0.0,0.0}
\definecolor{darkgreen}{rgb}{0.0,0.4,0.0}
\definecolor{darkblue}{rgb}{0.0,0.0,0.4}
\usepackage[bookmarks,linktocpage,colorlinks,
    linkcolor = darkred,
    urlcolor  = darkblue,
    citecolor = darkgreen]{hyperref}
\usepackage{subfigure}
\wocname{EPJ Web of Conferences}
\woctitle{Lattice2017}
\newcommand{\beq}{\begin{eqnarray}}
\newcommand{\eeq}{\end{eqnarray}}

\def\eq#1{Eq.~(\ref{#1})}
\def\spr{\!\cdot\!}

\def\beal{\begin{align}}
\def\eeal{\end{align}}
\def\bea{\begin{eqnarray}}
\def\eea{\end{eqnarray}}
\def\beq{\begin{equation}}
\def\eeq{\end{equation}}

\def\1eq#1{Eq.~(\ref{#1})}
\def\eq#1{Eq.~(\ref{#1})}

\def\2eqs#1#2{Eqs.~(\ref{#1}) and~(\ref{#2})}
\def\3eqs#1#2#3{Eqs.~(\ref{#1}),~(\ref{#2}) and~(\ref{#3})}

\def\fig#1{Fig.~\ref{#1}}

\def\ie{{\it i.e.}, }

\def\n#1{({\it #1})}

\def\t{\lambda}

\def\gl{A}
\def\Zgl{Z_\gl}

\def\ffT{T}

\def\ffGTR{\Gamma_{T\!,\,R}}

\def\ffTs{T^{\mathrm{sym}}}
\def\ffSs{S^{\!\mathrm{sym}}}
\def\ffGTs{\Gamma_T^{\mathrm{sym}}}
\def\ffGSs{\Gamma_S^{\mathrm{sym}}}

\def\ffTas{T^{\mathrm{asym}}}
\def\ffGTas{\Gamma_T^{\mathrm{asym}}}

\begin{document}
%
\selectlanguage{english}
\title{
On the zero-crossing of the three-gluon Green's function from lattice simulations}
\author{
\firstname{Andreas} \lastname{Athenodorou}\inst{1} \and
\firstname{Philippe} \lastname{Boucaud}\inst{2} \and
\firstname{Feliciano} \lastname{de Soto}\inst{3,8} \and
\firstname{Jos\'e} \lastname{Rodr\'{\i}guez-Quintero}\inst{4,8}\fnsep\thanks{Speaker, \email{jose.rodriguez@dfaie.uhu.es}} \and
\firstname{Savvas} \lastname{Zafeiropoulos}\inst{5,6,7} 
}
\institute{
Department of Physics, University of Cyprus, POB 20537, 1678 Nicosia, Cyprus
\and 
Laboratoire de Physique Th\'eorique (UMR8627), CNRS,
Univ. Paris-Sud, Universit\'e Paris-Saclay, 91405 Orsay, France
\and 
Dpto. Sistemas F\'{\i}sicos, Qu\'{\i}micos y Naturales, 
Univ. Pablo de Olavide, 41013 Sevilla; Spain
\and 
Dpto. Ciencias Integradas, Fac. Ciencias Experimentales; 
Universidad de Huelva, 21071 Huelva; Spain.
\and 
Thomas Jefferson National Accelerator Facility, Newport News, VA 23606, USA\and
Department of Physics, College of William and Mary, Williamsburg, VA 23187-8795, USA\and
Institute for Theoretical Physics, Universit\"at Heidelberg, Philosophenweg 12, D-69120 Germany
\and 
CAFPE, Universidad de Granada, E-18071 Granada, Spain
}
\abstract{
We report on some efforts recently made in order to gain a better understanding of some IR properties of the 3-point gluon Green's function by exploiting results from large-volume quenched lattice simulations. These lattice results have been obtained by using both tree-level Symanzik and the standard Wilson action, in the aim of assessing the possible impact of effects presumably resulting from a particular choice for the discretization of the action. The main resulting feature is the existence of a negative logaritmic divergence at zero-momentum, which pulls the 3-gluon form factors down at low momenta and, consequently, yields a zero-crossing at a given deep IR momentum. The results can be correctly explained by analyzing the relevant Dyson-Schwinger equations and appropriate truncation schemes.}
\maketitle

\section{Introduction}
\label{intro}

In the last few years, a very thorough scrutiny of the Quantum Chromodynamics (QCD) fundamental Green's functions using large-volume lattice simulations (see, for instance~\cite{Cucchieri:2010xr,Bogolubsky:2009dc,Oliveira:2009eh, Ayala:2012pb,Duarte:2016iko,Boucaud:2017ksi}, together with a variety of continuum approaches (see, for instance~\cite{Aguilar:2008xm,Boucaud:2008ky,Fischer:2008uz,Dudal:2008sp,Kondo:2011ab,Szczepaniak:2010fe,Watson:2010cn}), has taken place, aiming at a better understanding of the infrared (IR) sector of QCD. Notwithstanding that off-shell Green's functions are not physical quantities, given their explicit dependence on the 
gauge-fixing parameter and the renormalization scheme, they encode valuable information on fundamental  nonperturbative 
phenomena such as confinement, chiral symmetry breaking, and dynamical mass generation,
and constitute the basic building blocks of symmetry-preserving formalisms 
intended to provide with a faithful description of hadron phenomenology (see, for instance~\cite{Maris:2003vk,Chang:2011vu,Qin:2011xq,Eichmann:2012zz,Cloet:2013jya,Binosi:2014aea}).

A few recent papers~\cite{Athenodorou:2016oyh,Rodriguez-Quintero:2017phk,Boucaud:2017obn} made an effort to investigate further a key feature of the 3-point gluon Green function, the appearence of a zero-crossing at very low IR momentum caused by a negative logarithmic singularity at zero-momentum, by both exploiting up-to-date lattice data and accomodating these results within an alternative DSE approach. We will briefly review here these few works. 

\section{Connected and 1-PI 3-gluon Green's functions}
\label{sec-1}

Let us first properly define the connected and the usual 1-particle irreducible (1-PI) 3-gluon Green functions and describe then how they can be nonperturbatively obtained. 

\subsection{Definitions and generalities}

The connected three-gluon  vertex is defined as the correlation function~\footnote{A term proportional to the fully symmetric tensor, $d^{abc}$, can be only generally excluded in Landau gauge.} ($q+r+p=0$)
\begin{align}\label{eq:3gL}
	{\cal G}^{abc}_{\alpha\mu\nu}(q,r,p)=\langle{\gl^a_\alpha(q)}{\gl^b_\mu(r)}{\gl^c_\nu(p)}\rangle=f^{abc}{\cal G}_{\alpha\mu\nu}(q,r,p),
\end{align}
where the sub (super) indices represent Lorentz (color) indices and the average $\langle \cdot \rangle$ indicates functional integration over the gauge space. In terms of the 1-PI function, one has
\begin{align}
	{\cal G}_{\alpha\mu\nu}(q,r,p)&=g\Gamma_{\alpha'\mu'\nu'}(q,r,p)\Delta_{\alpha'\alpha}(q)\Delta_{\mu'\mu}(r)\Delta_{\nu'\nu}(p),
	\label{Cto1-PI}
\end{align}
with $g$ the strong coupling constant. In the Landau gauge, the transversality of the gluon propagator, {\it viz.}, 
\begin{align}\label{eq:propL}
	\Delta^{ab}_{\mu\nu}\left(q\right)=\langle \gl^a_\mu(q) \gl^b_\nu(-q) \rangle=  \delta^{ab} \Delta(p^2) P_{\mu\nu}(q),  
\end{align}
where $P_{\mu\nu}(q)=\delta_{\mu \nu} - q_\mu q_\nu/q^2$, implies directly that $\cal G$ is totally transverse: $q\spr{\cal G}=r\spr{\cal G}=p\spr{\cal G}=0$. 

\begin{figure*}[!t]
\begin{center}
	\hspace{-.75cm}
	\includegraphics[width=0.46\linewidth]{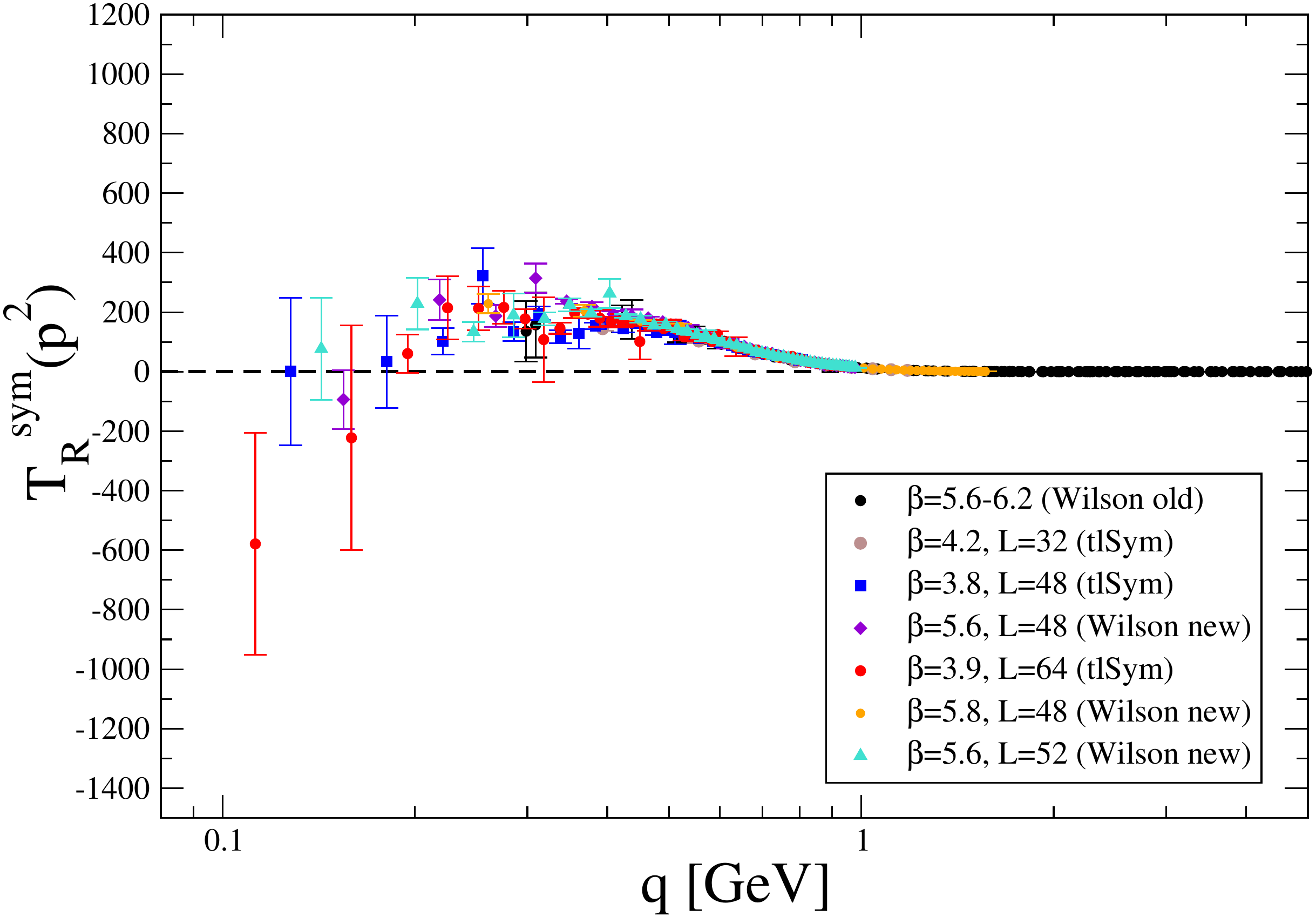}
	\includegraphics[width=0.46\linewidth]{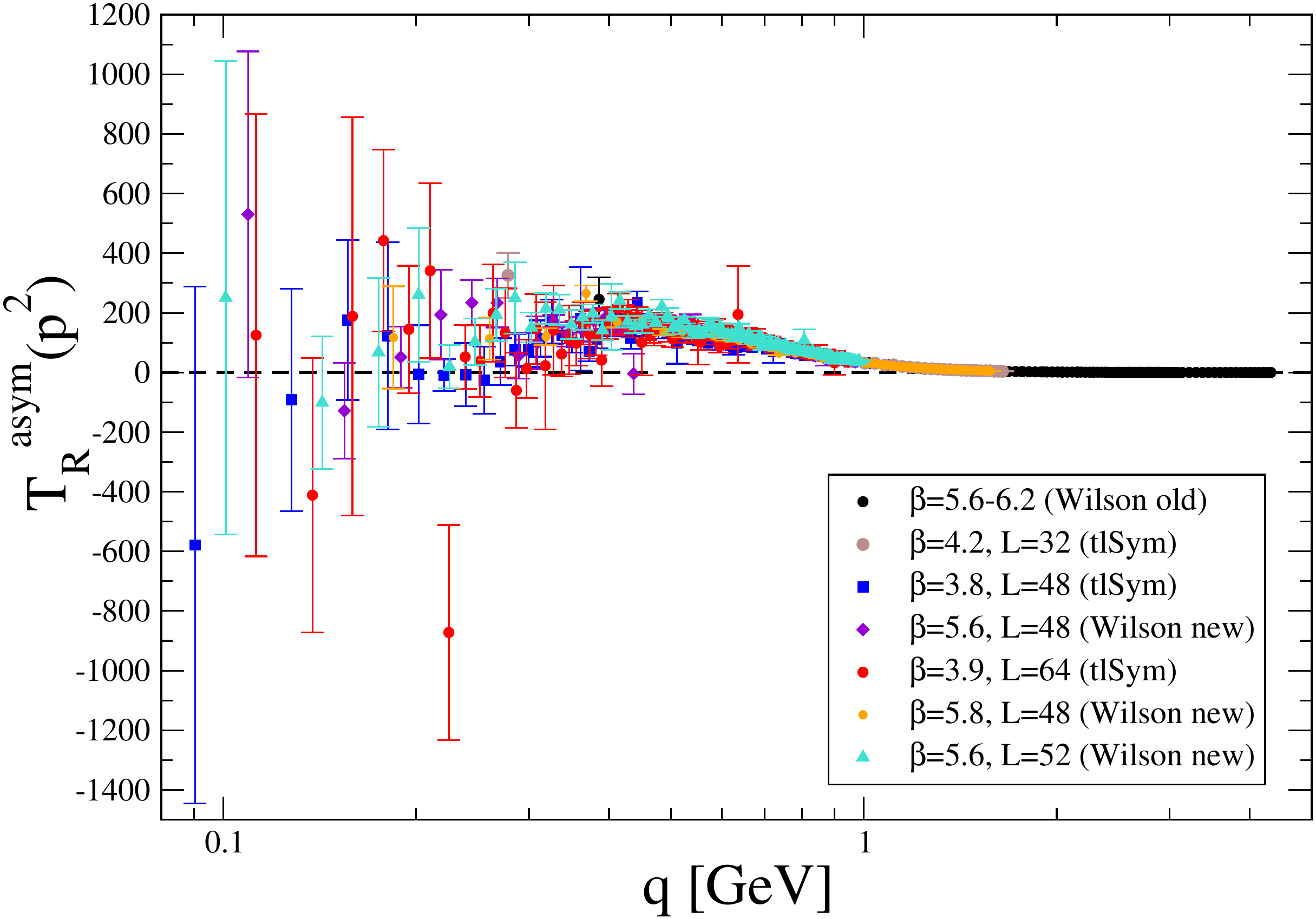}
\end{center}
	\caption{\label{latticeT}(color online) Lattice results for the renormalized connected form factor $\ffT_R$ in the symmetric (left) and asymmetric (right) momentum configuration. For both data sets the renormalization point $\mu=4.3$ GeV was chosen. The same scale is used in both plots which reveals the similar behavior of the two form factors.}
\end{figure*}

In what follows we will consider two special momenta configurations. The first one is the so-called symmetric configuration, in which $q^2=r^2=p^2$ and $q\spr r=q\spr p=r\spr p=-q^2/2$; in this case, there are only two totally transverse tensors, namely
\begin{align}
	\t^{\mathrm{tree}}_{\alpha\mu\nu}(q,r,p)&=\Gamma^{(0)}_{\alpha'\mu'\nu'}(q,r,p)
	 P_{\alpha'\alpha}(q) P_{\mu'\mu}(r) P_{\nu'\nu}(p),\nonumber \\
\t^{S}_{\alpha\mu\nu}(q,r,p)&=(r-p)_{\alpha}(p-q)_{\mu}(q-r)_{\nu}/r^2,
\end{align}  
where $\Gamma^{(0)}_{\alpha\mu\nu}$ is the usual tree-level vertex. Indicating with $\ffSs$ and $\ffTs$ (respectively, $\ffGSs$ and $\ffGTs$) the corresponding form factors in the decomposition of ${\cal G}$ (respectively, $\Gamma$) in this momentum configuration, \1eq{Cto1-PI} implies the relation
\begin{align}
	\ffTs(q^2)&=g\,\Gamma^{\mathrm{sym}}_{T}(q^2)\,\Delta^3(q^2), \nonumber \\
	\ffSs(q^2)&=g\,\Gamma^{\mathrm{sym}}_{S}(q^2)\,\Delta^3(q^2).
\label{Conn-1PI}
\end{align}
In particular, the $\ffTs$ form factor can be projected out through 
\begin{align}
	\ffTs(q^2)&=\left.\frac{W_{\alpha\mu\nu}(q,r,p)\,{\cal G}_{\alpha\mu\nu}(q,r,p)}{W_{\alpha\mu\nu}(q,r,p)W_{\alpha\mu\nu}(q,r,p)}\right\vert_\mathrm{sym},
	\label{prsym}
\end{align}
with $W=\t^{\mathrm{tree}}+\t^{S}/2$.

The second configuration we will study, which will be called `asymmetric' in what follows, is defined by taking the $q\to0$ limit, while imposing at the same time the condition $r^2=p^2=-p\spr r$. In this configuration $\t^{S}_{\alpha\mu\nu}\sim r_\alpha r_\mu r_\nu$ becomes totally longitudinal, and the only transverse tensor one can construct is obtained by the $q\to0$ limit of $\t^{\mathrm{tree}}$ (obviously omitting the $q$ projector), \ie
\begin{align}
	\t^{\mathrm{tree}}_{\alpha\mu\nu}(0,r,-r)&=2r_\alpha P_{\mu\nu}(r).
\end{align}
Thus one is left with a single form factor, which can be projected out through
\begin{align}
	\ffTas(r^2)&=\left.\frac{W_{\alpha\mu\nu}(q,r,p)\,{\cal G}_{\alpha\mu\nu}(q,r,p)}{W_{\alpha\mu\nu}(q,r,p)W_{\alpha\mu\nu}(q,r,p)}\right\vert_\mathrm{asym}\nonumber \\
	&=g\,\ffGTas(r^2)\,\Delta(0)\,\Delta^2(r^2),
	\label{prasym}
\end{align}
where now $W=\t^{\mathrm{tree}}$.

All the quantities defined so far are bare, and a dependence on the regularization cut-off must be implicitly understood. Within a given renormalization procedure, the renormalized Green's functions are calculated in terms of the renormalized fields $\gl_R=\Zgl^{-1/2}\gl$, so that
\begin{align}
	\Delta_R(q^2;\mu^2)&=\Zgl^{-1}(\mu^2)\,\Delta(q^2),\nonumber \\
	\ffTs_R(q^2;\mu^2)&=\Zgl^{-3/2}(\mu^2)\ffTs(q^2),
\end{align} 
and similarly for the asymmetric configuration. Within the MOM scheme that 
we will employ, one then requires that all the Green's functions take their tree-level expression at the subtraction point, 
namely
\begin{align}
	\Delta_R(q^2;q^2)&=\Zgl^{-1}(q^2)\,\Delta(q^2)=1/q^2,\nonumber \\
	\ffTs_R(q^2;q^2)&=\Zgl^{-3/2}(q^2)\,\ffTs(q^2)=g^{\mathrm{sym}}_R(q^2)/q^6.
\end{align}
The first equation yields the renormalization constant $\Zgl$ as a function of the bare propagator, which when substituted 
into the second equation provides a renormalization group invariant definition of the three-gluon MOM running coupling~\cite{Alles:1996ka,Boucaud:1998bq}:
\begin{align}
	g^{\mathrm{sym}}(q^2)&=q^3\frac{\ffTs(q^2)}{[\Delta(q^2)]^{3/2}}=q^3\frac{\ffTs_R(q^2;\mu^2)}{[\Delta_R(q^2;\mu^2)]^{3/2}}.
	\label{alpha3gsym}
\end{align}
In the asymmetric configuration the relation is slightly different, as in this case one has
\begin{align}
	\ffTas_R(r^2;r^2)&=\Zgl^{-3/2}(r^2)\,\ffTas(r^2)=\Delta_R(0;q^2)\,g^{\mathrm{asym}}_R(r^2)/r^4,
\end{align}
implying 
\begin{align}
	g^{\mathrm{asym}}(r^2)&=r^3\frac{\ffTas(r^2)}{[\Delta(r^2)]^{1/2}\Delta(0)}=r^3\frac{\ffTas_R(r^2;\mu^2)}{[\Delta_R(r^2;\mu^2)]^{1/2}\Delta_R(0;\mu^2)}.
	\label{alpha3gasym}
\end{align}
Finally, in both cases the above equations yield for the 1-PI form factors the relation
\begin{align}\label{eq:1PIfromG}
	g^i(\mu^2)\,\ffGTR^i(\ell^2;\mu^2)&=\frac{g^i_R(\ell^2)}{[\ell^2\Delta(\ell^2;\mu^2)]^{3/2}},
\end{align}
where $i$ indicates either the symmetric or the asymmetric momentum configuration, and, correspondingly, $\ell^2=q^2,\,r^2$. 

This latter result is of special interest because it establishes a connection between the three-gluon MOM running coupling, which many lattice and continuum studies have paid attention to, and the vertex function of the amputated three-gluon Green's function, a fundamental ingredient within the tower of (truncated) SDEs addressing non-perturbative QCD phenomena. In fact, these quantities are related only by the gluon propagator $\Delta$, which, after the intensive studies of the past decade, is very well understood and accurately known.

\subsection{Lattice QCD results} 

The purpose of obtaining a nonperturbative estimate of the 1-PI and connected 3-gluon form factors will be achieved by the determination of the matrix elements defined in \2eqs{eq:3gL}{eq:propL} as, respectively, 2- and 3-points correlation functions of the gluon gauge fields obtained from quenched lattice simulations. The form factors can be related to the matrix elements by \2eqs{prsym}{prasym}, as it results from applying the appropriate projections described in the previous subsection. In the aim of concluding about qualitative features for the deep IR behavior of these form factors, we have simulated lattice volumes in physical units as large as possible but within the quenched approximation, under the working assumption that the light dynamical quarks only affect quantitatively this behavior. In particular, we have exploited quenched SU(3) gauge-field configurations at several large volumes and different bare couplings $\beta$, obtained employing both the tree-level Symanzik: 420 configurations at $\beta=4.2$ for a hypercubic lattice of length $L=32$ (corresponding to a physical volume of 4.5$^4$ fm$^4$), 2000 configurations at $\beta=3.90$ for $L=64$ lattice (15.6$^4$ fm$^4$) and 1050 configurations at $\beta=3.8$ for $L=48$ (13.7$^4$ fm$^4$); and the Wilson action: 960 configurations at $\beta=5.8$ for $L=48$ (6.72$^4$ fm$^4$), 1920 configurations at $\beta=5.6$ for $L=48$ (11.3$^4$ fm$^4$) and 1790 $\beta=5.6$ for $L=52$ (12.3$^4$ fm$^4$). These last data have been supplemented with those derived from the old configurations of~\cite{Boucaud:2002fx}, obtained using the Wilson gauge action at several $\beta$ (ranging from 5.6 to 6.0), lattices (from $L=24$ to $L=32$) and physical volumes (from 2.4$^4$ to 5.9$^4$ fm$^4$). 

The results can be found in~\fig{latticeT}, where we plot the form factor $\ffT$ renormalized at $\mu=4.3$ GeV for both the symmetric (left panel) and asymmetric (right panel) momentum configurations. In the symmetric case $\ffTs_R$ displays a zero crossing located in the IR region around 0.1--0.2 GeV, after which the data seems to indicate that some sort of divergent behavior manifests itself. In the asymmetric case the situation looks less clear as data are noisier, as it results from studying a correlation function where one of the fields is taken at zero momentum.

\section{Analysis of results}

We will now briefly describe the analysis of the results, already presented (including many more details) in ref.~\cite{Athenodorou:2016oyh}, mainly addressed to understand a  striking feature taking place in the low-momentum domain: the appearance of a zero-crossing and a negative logarithmic singularity at zero-momentum (many independent analyses within the SDE formalism, employing a variety of techniques and truncation schemes, have found the same in the 3-point~\cite{Aguilar:2013vaa,Blum:2014gna,Eichmann:2014xya,Cyrol:2016tym} and the 4-point~\cite{Binosi:2014kka,Cyrol:2014kca} gluon sector, and also when light quarks are included~\cite{poster}), the underlying origin of this phenomenon is the masslessness of the ghost propagators circulating in the {\it nonperturbative} ghost loop diagram contributing to the SDE of n-point Green's functions~\cite{Aguilar:2013vaa}. Specifically, employing a nonperturbative Ansatz for the gluon-ghost vertex that satisfies the correct STI, the leading IR contribution from the ghost-loop, denoted by $\Pi_c(q^2)$, is given by~\cite{Aguilar:2013vaa}   
\begin{align}
\Pi_c(q^2) = \frac{g^2 C_A}{6} q^2 F(q^2) \int_{k}\frac{F(k^2)}{k^2 (k+q)^2} \,,
\end{align}
where  $C_A$ is the Casimir eigenvalue in the adjoint representation, and  $\int_{k}\equiv {\mu^{\epsilon}}/{(2\pi)^{d}}\!\int\!\mathrm{d}^d k$ is the dimensional regularization measure, with $d=4-\epsilon$ and $\mu$ is the 't Hooft mass; evidently, in the limit $q^2\to 0$, the above expressions behave like $q^2 \log {q^2}/{\mu^2}$.  Even though this particular term does not interfere with the finiteness of  $\Delta(q^2)$, its presence induces two main effects: \n{i}  $\Delta(q^2)$ displays a mild maximum at some relatively low value of $q^2$, and  \n{ii} the first derivative of $\Delta^{-1}(q^2)$ diverges logarithmically at $q^2=0$. The form of the renormalized gluon propagator that emerges from the complete SDE analysis may be accurately parametrized in the IR by the expression  
\begin{align}
	\Delta_R^{-1}(q^2;\mu^2)\underset{q^2\to0}{=}q^2\left[a+b\log\frac{q^2+m^2}{\mu^2}+c\log \frac{q^2}{\mu^2} 
	\right]+m^2,
	\label{modindDelta}
\end{align}   
with $a$, $b$, $c$, and $m^2$ suitable parameters, which captures explicitly the two aforementioned effects. Note that  
\mbox{$\Delta_R^{-1}(0;\mu^2) =m^2$}, and that the `protected' logarithms stem from gluonic loops.  

\begin{figure*}[!t]
	\begin{center}
	\hspace{-.75cm}
	\includegraphics[width=0.46\linewidth]{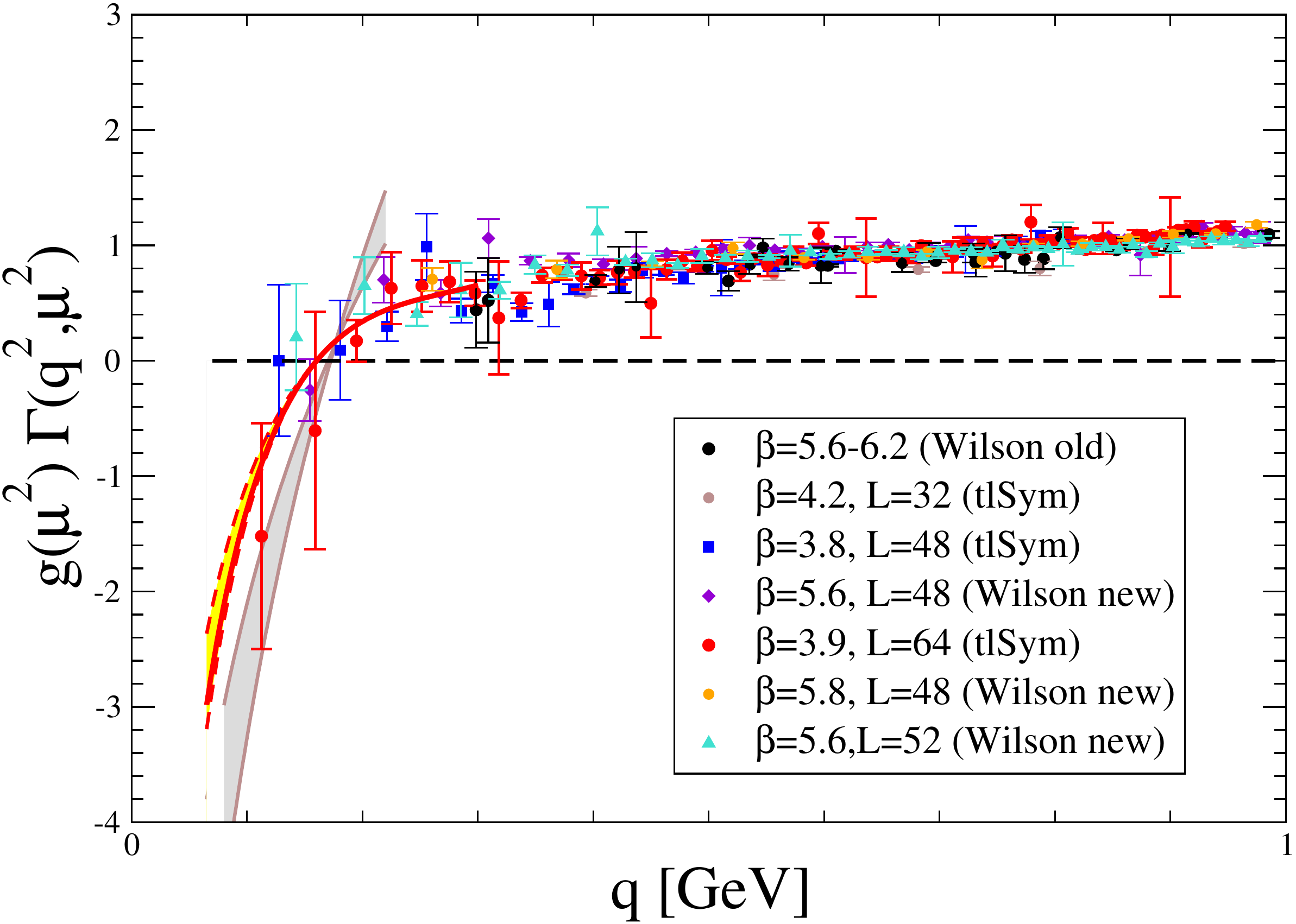}
	\includegraphics[width=0.46\linewidth]{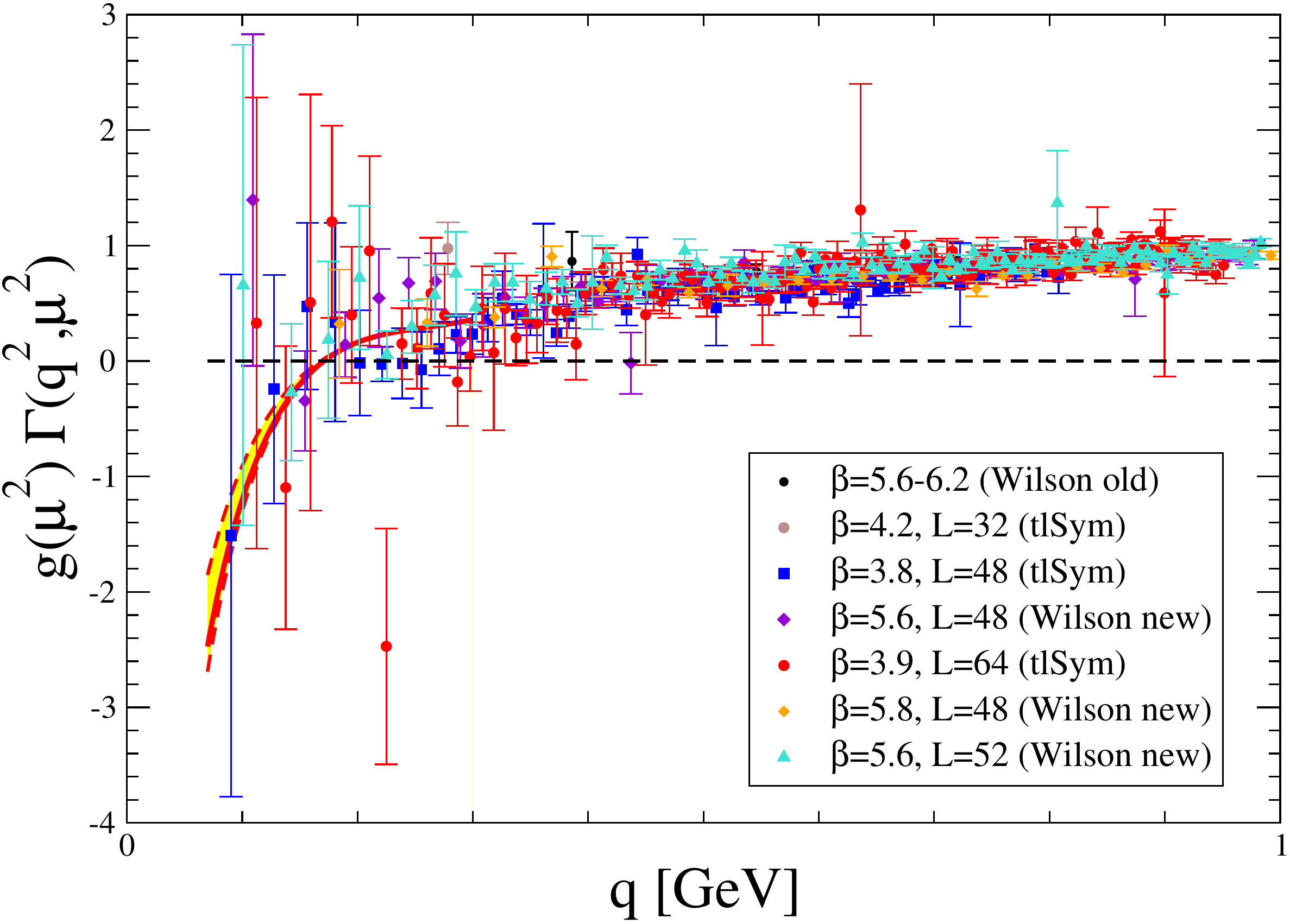}	
	\end{center}
	\caption{\label{compT} The 3-gluon 1-PI form factors, related to the connected Green's functions through~\2eqs{prsym}{prasym}, obtained from the lattice data displayed in Fig.~\ref{latticeT} although plotted now making use of a momentum linear scale, for symmetric (left) and asymmetric (right) momenta configuration.  The renormalization point is $\mu$=4.3 GeV. The red solid and dashed lines result from the best fits with \eq{eq:gGammaF}, while the brown solid lines correspond to fits done after dropping the subleading term driven by $a^i_2$ . Yellow and brown bands depict the uncertainty resulting from the the gauge two-point Green's function in the determination of $a^i_{ln}$~\cite{Athenodorou:2016gsa} for the fits.}
\end{figure*}

Any standard Green's function can be related to the same one with {\it background} legs, within the PT-BFM approach, by the use of the so-called ``{\it background quantum}" identities~\cite{Grassi:1999tp,Binosi:2002ft,Binosi:2009qm}). The ones with {\it background legs}, when projected according to Eqs.~(\ref{prsym},\ref{prasym}) and by virtue of the Abelian STI that the PT-BFM propagators are constructed to obey, will be led in the IR by the derivative of the inverse of the gluon propagator, represented by \eq{modindDelta}~\cite{Aguilar:2013vaa}. Thus, the three-gluon 1-PI form factors derived from the background Green's functions, $\Gamma^{(B)}$, can be proven to behave in the deep IR as
\begin{eqnarray}
\Gamma^{i,(B)}_{T,R}(p^2;\mu^2) &\underset{p^2/\mu^2\ll 1}{\simeq}&  F_R(0;\mu^2) \frac{\partial}{\partial p^2}  \Delta^{-1}_R(p^2;\mu^2) \ + \ \dots \nonumber \\ 
& \simeq & F_R(0;\mu^2) \left( a + b\ln\frac{m^2}{\mu^2} + c \right) + c \  F_R(0;\mu^2) \ln \frac{p^2}{\mu^2}  \ + \ \dots \ ;  
\label{eq:GammaB}
\end{eqnarray}
where $F_R(0;\mu^2)$ is the renormalized ghost dressing function evaluated at zero momentum and where the dots stand for subleading corrections that, as discussed in \cite{Athenodorou:2016gsa}, might be collectively taken into account by adding an extra constant term which, contrarily to the leading contribution, depends a priori on the momenta configuration. On the other hand, the connection between the background and the standard vertex functions, and that of $\Gamma^{(B)}$ and $\Gamma$, is controlled by the ghost-gluon dynamics and will essentially introduce subleading corrections, as it is also done by the low-momentum expansion of the ghost dressing function in \eq{eq:GammaB}, which do not modify the leading logarithmic divergence shown by this equation~\cite{Boucaud:2010gr,RodriguezQuintero:2011vw,Binosi:2016xxu}. Thus, aiming at a reliable description of the lattice data for the vertex 1-PI form factors within the IR domain, we can eventually write
\begin{equation}\label{eq:gGammaF}
g_R^i(\mu^2) \Gamma^i_R(p^2;\mu^2) \ = \ a_{ln}^i(\mu^2) \ln \frac{p^2}{\mu^2} \ + \ a_0^i(\mu^2) \ + \ 
a_2^i(\mu^2) \ p^2 \ln\frac{p^2}{M^2} \ + \ o(p^2) \ ,
\end{equation}
where $a_0^i$, $a_2^i$ and $M$ will be free parameters capturing subleading contributions, while $a_{ln}^i(\mu^2)=g_R^i(\mu^2) \ c \ F_R(0;\mu^2)$, is known from gluon and ghost two-point Green's functions and from the value of the three-gluon coupling at the renormalization point. $M$ differs a priori from the renormalization point since it is absorbing the ${\cal O}(p^2)$-contribution which, for the sake of consistency, is also required. Then, \eq{eq:1PIfromG} can be invoked to derive the estimates for the 1-PI form factors from the lattice data displayed in Fig.~\ref{latticeT}, and \eq{eq:gGammaF} applied to account for their IR behavior, where the zero-crossing feature takes place. A correct description of the IR behavior for these form factors makes thus also possible a reliable determination of the momentum for which the zero value is taken. The lattice data and the best fits to them with \eq{eq:gGammaF} (in solid red line) appear displayed in the right and left panels of Fig.~\ref{compT}, respectively, for symmetric and asymmetric renormalization schemes. Although it is evident that in the symmetric case a better description of the IR data is achieved, both cases happen to be remarkably consistent with each other and with the existence of a zero-crossing, as predicted by employing continuum nonperturbative approaches such as DSE. 

\section{Conclusions}

We have briefly reviewed some recent studies for the 3-gluon Green's function, renormalized by applying two different renormalization schemes,  obtained by exploiting large-volume lattice simulations without dynamical quarks. The working hypothesis underlying the reliability of the quenched approximation for our purpose is that the leading IR behavior of the 3-gluon form factor is only dominated by the divergent ghost loops entering in the gluon vacuum polarization through its DSE, only mildly affected at a quantitative level by the presence of light quarks. This assumption has been consistently supported by recent DSE anlysis~\cite{Williams:2015cvx}. The most striking feature of the three-gluon Green function, taking place in its very deep IR domain, which in particular is elusive to the semiclassical description of ref.~\cite{Athenodorou:2016gsa,Athenodorou:2018jwu}, is the existence of a negative logarithmic singularity at zero momentum which causes the appearance of a zero-crossing, owing to the masslessness of the ghost which contributes via nonperturbative ghost-loops to the SDE of the gluon Green's functions. This is an important phenomenon, with dynamical implications~\cite{Aguilar:2013vaa,Binosi:2016xxu} which, notwithstanding that it takes place within the deep IR momentum domain, can be hardly captured within the framework of a semiclassical approach.

\section{Acknowledgements}
 {\small This work has been partially funded by the Spanish Ministry research project FPA2014-53631-C2-2-P.  
SZ acknowledges support by the National Science Foundation (USA) under
grant PHY-1516509 and by the Jefferson Science Associates,
LLC under U.S. DOE Contract \# DE-AC05-
06OR23177. SZ is also indebted to A. Sciarra
for all his help regarding the CL2QCD code.  CL2QCD is a Lattice QCD application based on OpenCL, applicable
to CPUs and GPUs.  Numerical computations have used resources of CINES and GENCI-IDRIS as well as resources
at the IN2P3 computing facility in Lyon}


\end{document}